\documentclass[preprint,prb,aps,amsmath,amssymb]{revtex4-1}

\usepackage{graphicx}

\begin{document}

\title{Comparative study on magnetoresistance of carbon-cobalt nanocomposite thin films grown by pulsed laser deposition}

\author{S. Sergeenkov$^{1,*}$, C. C\'{o}rdova$^{2}$, L. Cichetto Jr$^{2,3,4}$,  O.F. de Lima$^{5}$, E. Longo$^{3,4}$,  F.M. Ara\'{u}jo-Moreira$^{2}$, C. Furtado$^{1}$}

\affiliation{$^{1}$\textit{Department of Physics, Universidade Federal da Para\'{i}ba,  Jo$\tilde{a}$o Pessoa, PB, Brazil}\\ $^{2}$\textit{Department of Physics, Universidade Federal de S$\tilde{a}$o Carlos, S$\tilde{a}$o Carlos, SP,  Brazil}\\ $^{3}$\textit{LIEC, Department of Chemistry, Universidade Federal de S$\tilde{a}$o Carlos,  S$\tilde{a}$o Carlos, SP, Brazil}\\ $^{4}$\textit{Institute of Chemistry, Universidade Estadual Paulista - UNESP, Araraquara, SP, Brazil}\\ $^{5}$\textit{Institute of Physics 'Gleb Wataghin', UNICAMP, Campinas, SP, Brazil}\\
$^{*}$Corresponding author; email: sergei@df.ufscar.br}

\date{\today}

\begin{abstract}
We present a comparative study on the influence of applied magnetic field on the resistance of $C_{1-x}Co_x$ thin films (with $x=0.1$, $0.15$ and $0.2$) grown on $Si$ substrate by pulsed laser deposition technique. It is found that the behavior of magnetoresistance (MR) drastically depends on the temperature. Namely, at low temperatures MR is positive and its behavior is governed by the field  mediated weak localization scenario. While at high temperatures MR turns negative and its behavior is dominated by electron scattering on ferromagnetic cobalt atoms.
\end{abstract}

\maketitle

\section{Introduction}

As is well known, \cite{1,2,3,4,5,6,7,8} spintronics is based on a possibility to control electric properties with applied magnetic fields. Recent progress in numerous applications of nanomaterials \cite{9,10,11,12,13,14,15,16,17,18,19} has proved nanostructured doped carbon thin films one of the most promising materials. Needless to say that the performance of the devices based on these materials essentially depends on the clear understanding of the underlying basic mechanisms which govern their structural, magnetic and transport properties.  Of particular interest is manifestation of the spin-dependent transport properties of magnetically ordered metallic atoms implanted into non-magnetic carbon matrix. \cite{20,21,22,23,24,25} It has been well established that at low temperature, the magnetoresistance (MR) $\Delta R(T,H)=R(T,H)-R(T,0)$ can change from positive (in small fields) to negative (in larger fields) behavior due to the change from a weak localization (spin polaron tunneling) \cite{2} to antilocalization (spin reorientation upon electron scattering). \cite{3}  Likewise, at high temperatures, the spin-dependent scattering of conduction electrons on ferromagnetic Co nanograins can be responsible both for positive and negative MR \cite{7,8}  depending on the relationship between $\Delta R(T,H)$ and magnetization $M(T,H)$. 

In this paper we present our latest results on magneto-transport properties of $C_{1-x}Co_x$  thin films (with $0.1<x<0.2$) grown by the Pulsed Laser Deposition (PLD) technique. \cite{26,27,28,29,30,31} Some interesting features in the magnetic field and doping dependences of the measured magnetoresistance have been observed and their origin has been discussed in detail using the up-to-date transport scenarios.

\section{Experimental methods}

High quality $C_{1-x}Co_x$ thin films with $x=0.10$, $0.15$ and $0.20$ have been grown by PLD technique. Graphite and cobalt superfine powders of high purity ($99.98\%$) were used. 
Thin films have been deposited on $Si(100)$ ($5 \times 5 mm^{2}$) substrates, using a $KrF$ excimer laser (with wavelength $\lambda = 248nm$ and $25ns$ pulse width). Microstructure and crystallographic orientation of the films were characterized by X-ray diffraction scans. The films thickness ($85.3nm$) was confirmed by using field-emission scanning electron microscopy (FEG SEM). Further details on preparation and characterization of our films can be found elsewhere. \cite{29} 
The zero-field electrical resistance was measured using the conventional four-probe method. To avoid Joule and Peltier effects, a dc current $I=100\mu A$ was injected (as a one second pulse) successively on both sides of the sample. The voltage drop $V$ across the sample was measured with high accuracy by a $KT256$ nanovoltmeter. Magnetic measurements were performed using a Quantum Design-MPMS-5 SQUID magnetometer working in the temperature range from $2$ to $400K$ and the field range from $0$ to $50kOe$.

\section{Results and discussion}

\subsection{Low-temperature magnetoresistance}

A careful analysis of the obtained experimental data revealed that $R(T,H)$ follows markedly different behavior at low and high temperatures, which can be well accounted for assuming the existence of two major scattering mechanisms (see below). 
The typical results for the low-temperature ($T=2.5K$) magnetic field dependence of $R(T,H)$ along with the best fits (solid lines) are shown in Fig.1. 

\begin{figure}
\centerline{\includegraphics[width=5.50cm]{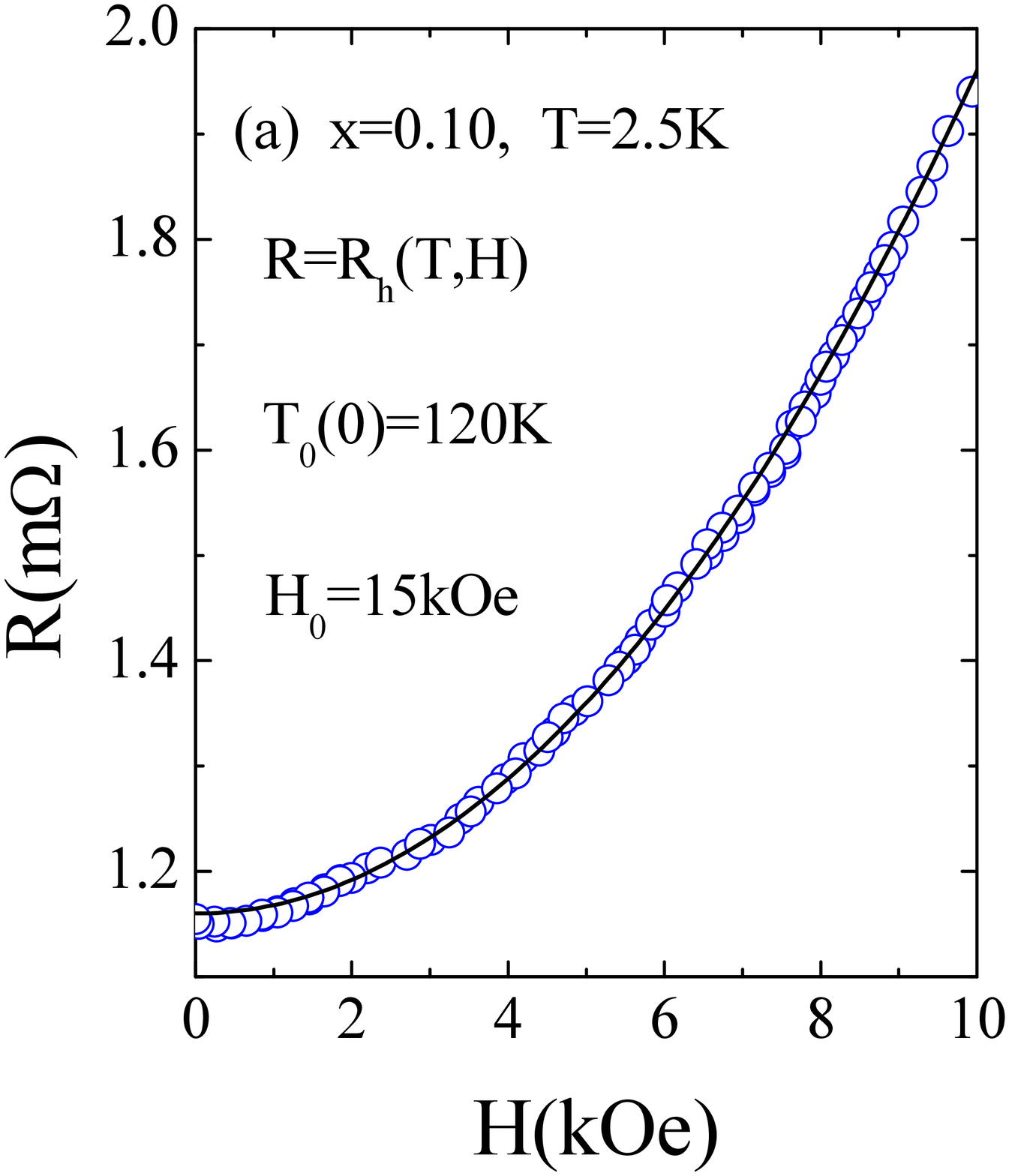} }\vspace{0.5cm}
\centerline{\includegraphics[width=5.50cm]{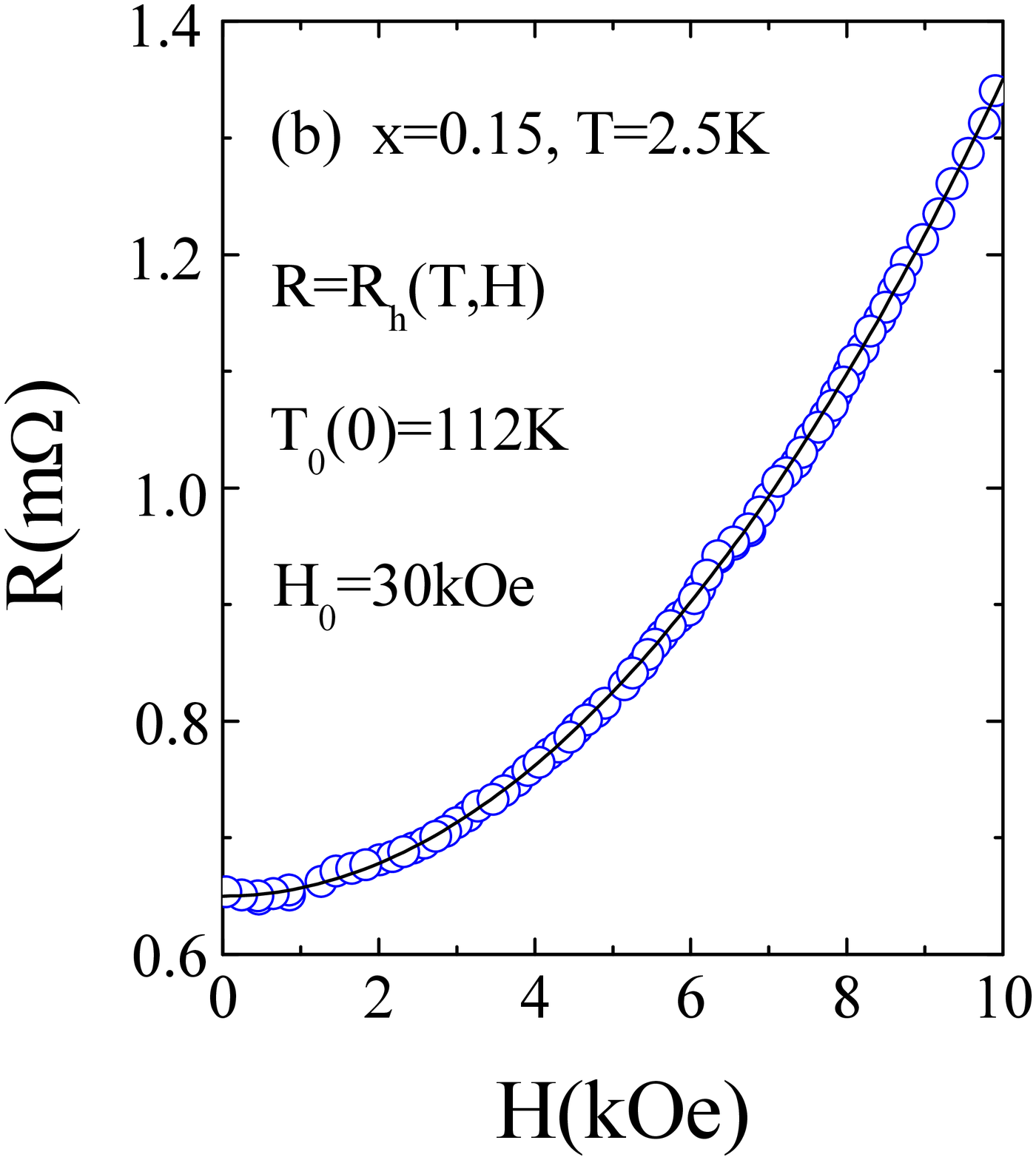} }\vspace{0.5cm}
\centerline{\includegraphics[width=5.50cm]{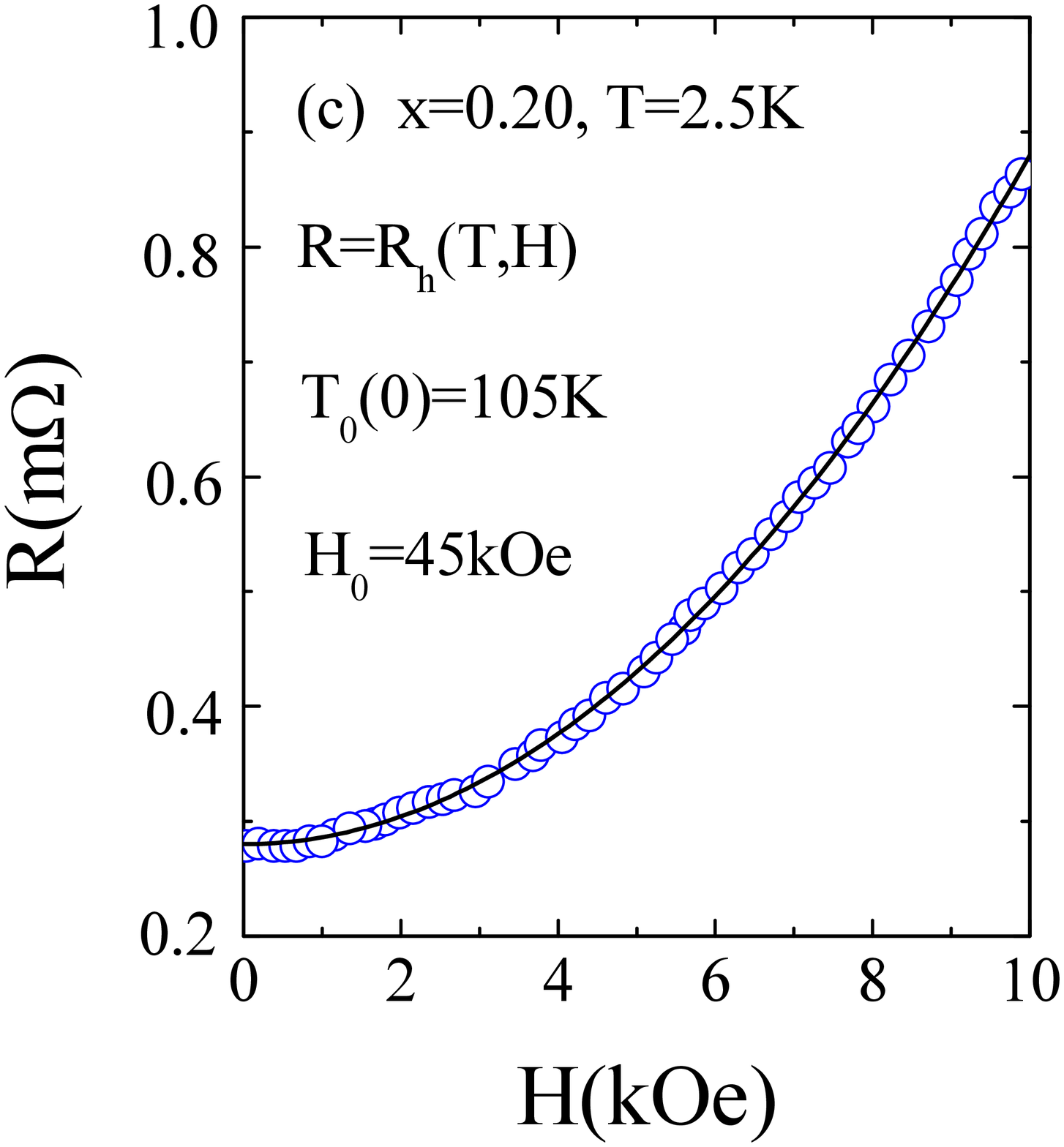} }\vspace{0.1cm}
\caption{Low-temperature ($T=2.5K$) magnetic field dependence of the resistance $R(T,H)$ measured for $C_{1-x}Co_x$  thin films with different content of $Co$ atoms: (a) $x=0.1$, (b) $x=0.15$, and (c) $x=0.2$. The solid lines are the best fits for hopping $R_h(T,H)$ contribution according to Eq.(1).}
\label{fig:fig1}
\end{figure}

Based on our previous zero-field resistance results, \cite{29} it is reasonable to assume that in the low-temperature ferromagnetic (FM) region $R(T,H)$ is dominated by a spin-polaron hopping scenario modified by applied magnetic field, leading to the following expression \cite{29,32,33,34,35}

\begin{equation}
R_h(T,H)= R_{oh}(x)\exp \left \{\left[\frac{T_0(H,x)}{T}\right]^{0.5} \right \}  
\end{equation}
with a localization temperature 
\begin{equation}
T_0(H,x)= T_0(0,x)\left \{1+\left[\frac{H}{H_0(x)}\right]^2 \right \}
\end{equation}
Here $H_0(x)$ is the localization field.

Recall \cite{33,34} that the localization temperature $T_0(H,x)\simeq E_{\sigma }(H,x)/k_B$ depends on the field modulated spin-polaron tunneling energy $E_{\sigma }(H,x)$ related to the charge carrier localization length $L(H,x)\simeq \hbar /\sqrt{2mE_{\sigma }(H,x)}\simeq L(0,x)/\sqrt{1+H^2/H_0^2(x)}$ with $L(0,x)\simeq L(0,0)/\sqrt{1-x}$ (here $m$ is an effective carrier mass). According to Fig.2, the extrapolated zero-field localization temperature follows a linear $x$ dependence, $T_0(0,x)\simeq T_0(0,0)(1-x)$ with $T_0(0,0)\simeq 133K$, while the corresponding field shows percolation-like behavior $H_0(x)\simeq H_0(x_m)(x/x_m)^{\alpha}$ with $H_0(x_m)=45kOe$,  $\alpha =1.5$, and $x_m=0.2$. 

In turn, all these dependences produce $L(0,x_m)\simeq 2.2nm$ for the variation of the zero-field charge carrier localization length (assuming a free electron mass value for $m$). Notice that at low temperature, the MR $\Delta R_h(T,H)=R_h(T,H)-R_h(T,0)$ is positive. Since $L(H,x)<L(0,x)$, such a behavior can indeed be associated with a weak localization scenario at small magnetic fields. \cite{2}  On the other hand, at larger fields, $\Delta R_h(T,H)$ turns negative due to spin reorientation upon electron scattering which results in the antilocalization scenario. \cite{3}

\begin{figure}
\centerline{\includegraphics[width=6.0cm]{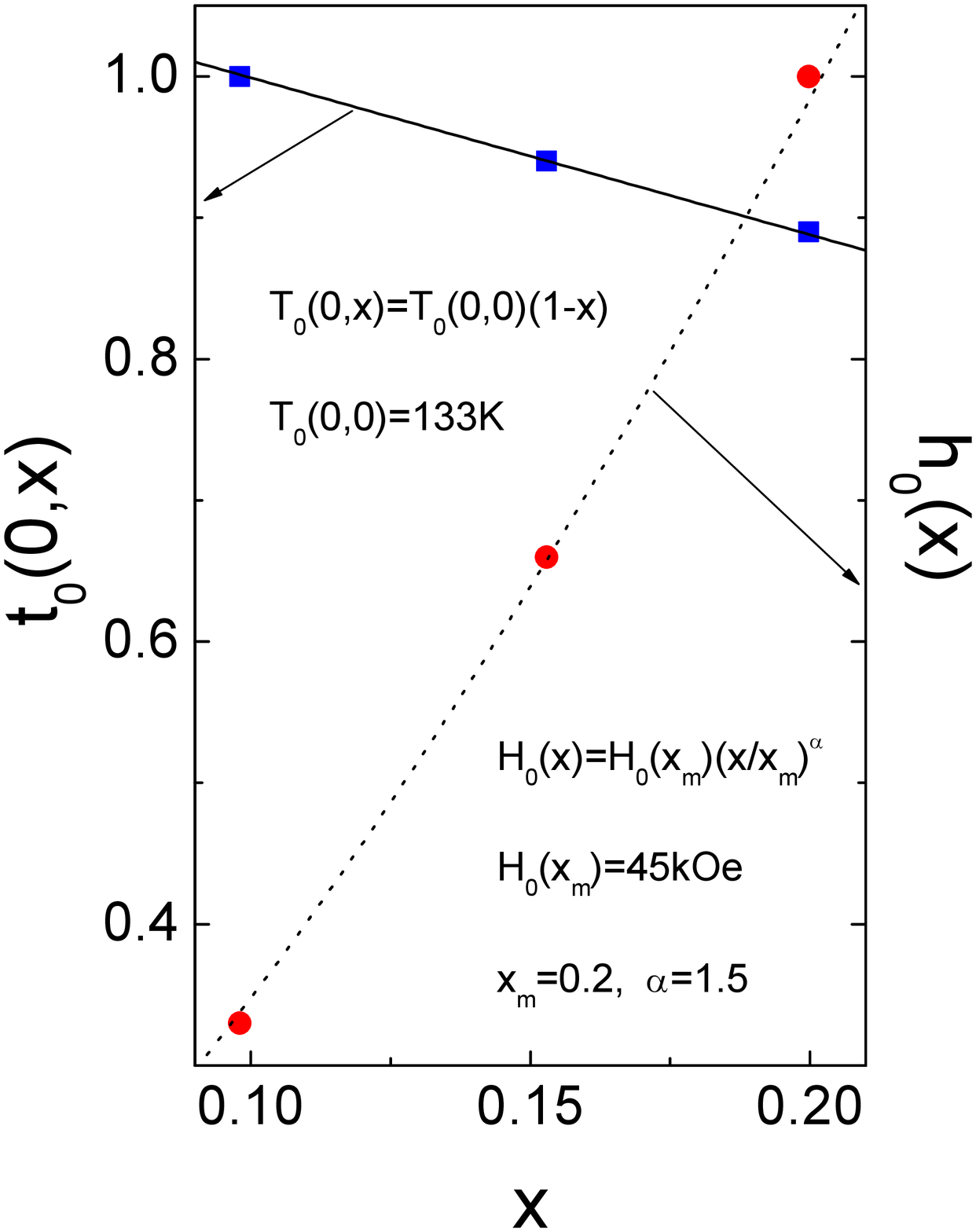} }\vspace{0.1cm}
\caption{The extracted concentration dependence of normalized zero-field localization temperature $t_0(x)=T_0(0,x)/T_0(0,0)$ and the corresponding intrinsic field $h_0(x)=H_0(x)/H_0(x_m)$.}
\label{fig:fig2}
\end{figure}

\subsection{High-temperature magnetoresistance}

\begin{figure}
\centerline{\includegraphics[width=5.50cm]{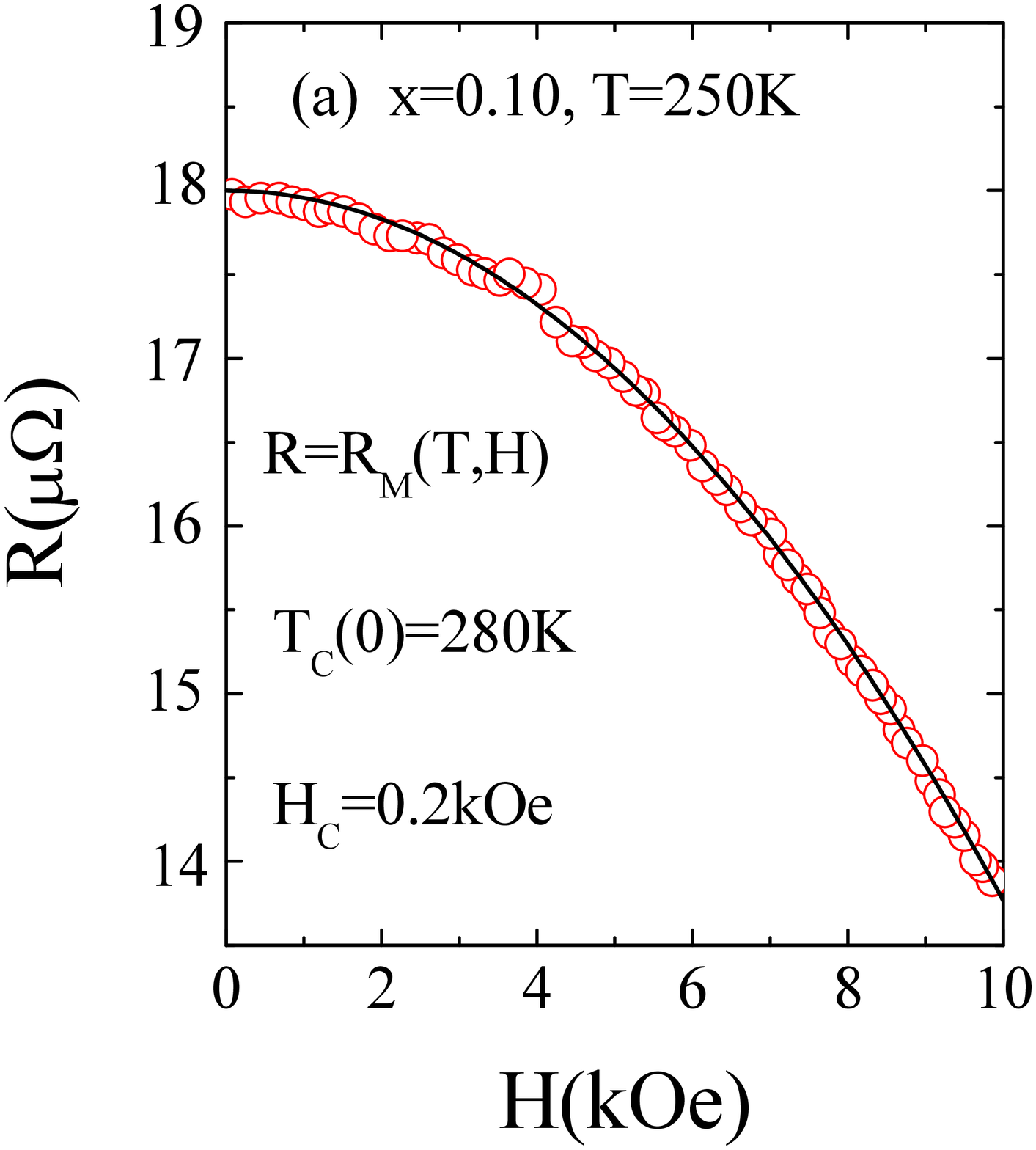} }\vspace{0.50cm}
\centerline{\includegraphics[width=5.50cm]{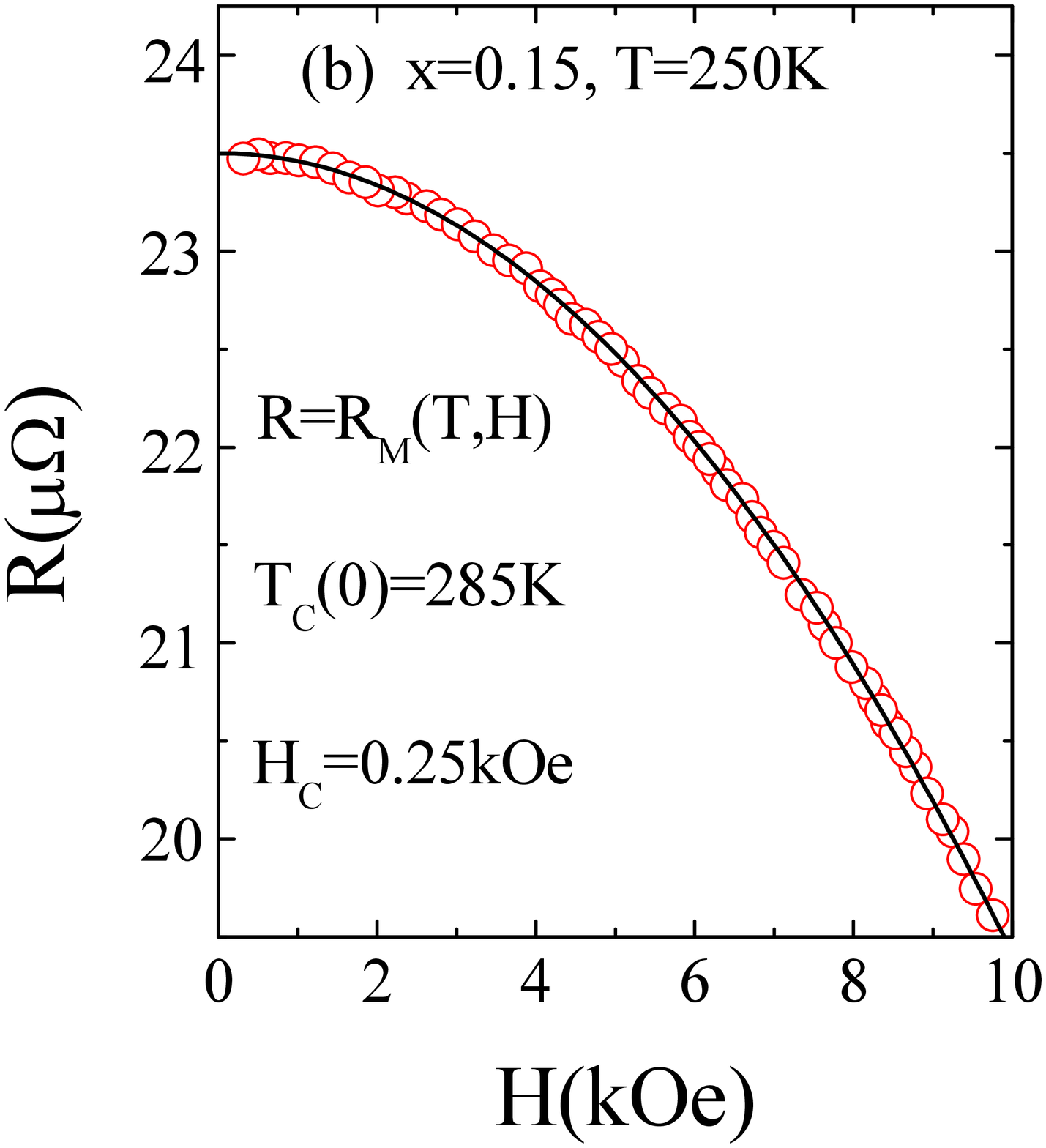} }\vspace{0.50cm}
\centerline{\includegraphics[width=5.50cm]{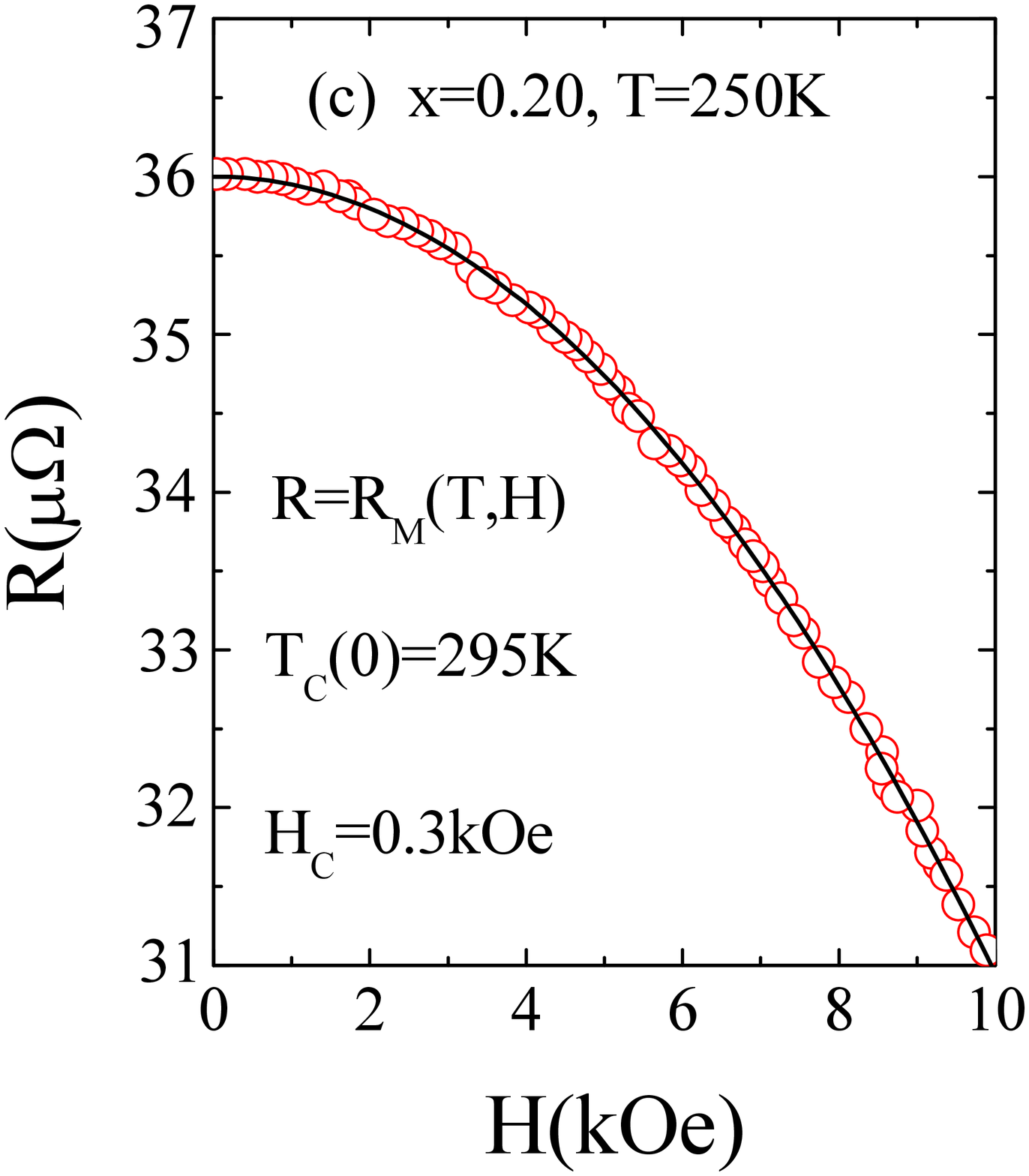} }\vspace{0.1cm}
\caption{High-temperature ($T=250K$) magnetic field dependence of the resistance $R(T,H)$ measured for $C_{1-x}Co_x$  thin films with different content of $Co$ atoms: (a) $x=0.1$, (b) $x=0.15$, and (c) $x=0.2$. The solid lines are the best fits for magnetic scattering $R_M(T,H)$ contribution according to Eq.(3).}
\label{fig:fig3}
\end{figure}
At the same time, according to Fig.3, which shows the high-temperature region (at $T=250K$), the magnetoresistance seems to be dominated by electron scattering on cobalt created FM structure with field modified magnetization $M(T,H)$, namely \cite{29,35}
\begin{equation}
R_M(T,H)=q M(T,H)
\end{equation}
with
\begin{equation}
M(T,H)= M_s(x)\tanh\left\{\left[\frac{T_C(H,x)}{T}\right]^{4}-1 \right \} 
\end{equation}
where 
\begin{equation}
T_C(H,x)= T_C(0,x)\left \{1-\left[\frac{H}{H_c(x)}\right]^2 \right \}
\end{equation}
is the field dependent Curie temperature and $M_s(x)\propto 1/H_c(x)$ is the saturation magnetization with $H_c(x)$ being the corresponding coercive field.

\begin{figure}
\centerline{\includegraphics[width=6.0cm]{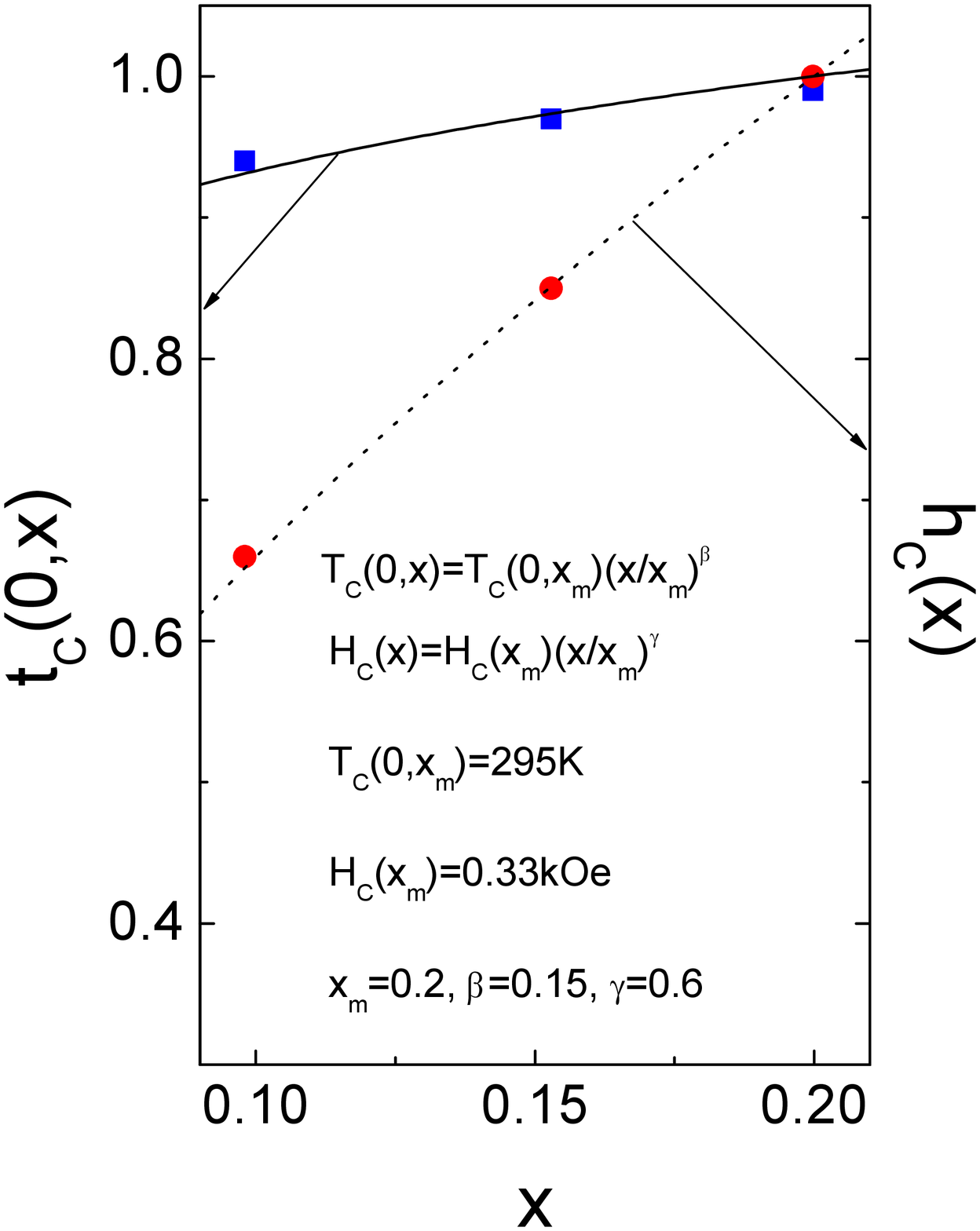} }\vspace{0.1cm}
\caption{The extracted concentration dependence of normalized zero-field Curie temperature $t_C(0,x)=T_C(0,x)/T_C(0,x_m)$ and the coercive field $h_c(x)=H_c(x)/H_c(x_m)$.}
\label{fig:fig4}
\end{figure}

\begin{figure}
\centerline{\includegraphics[width=8.50cm]{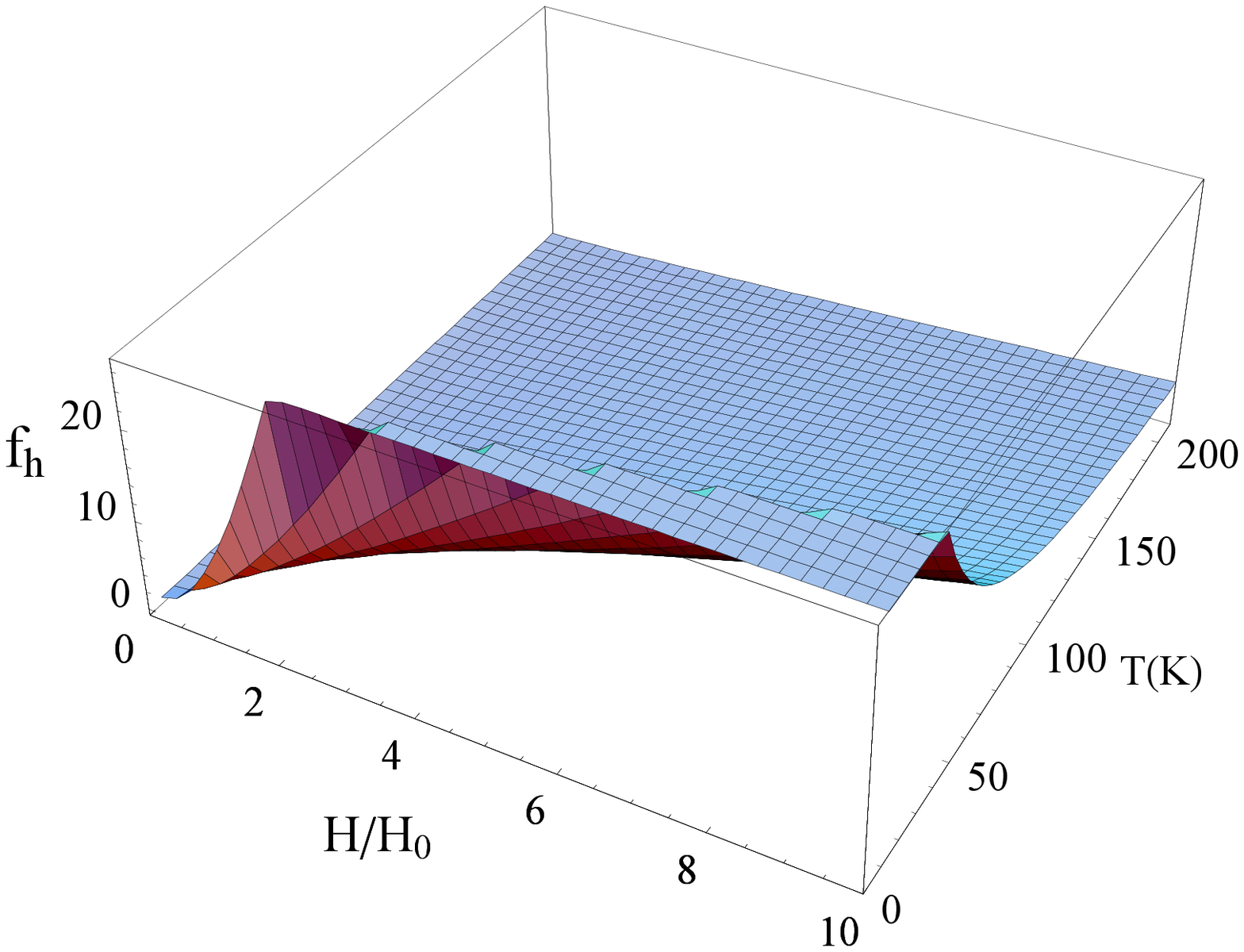}}\vspace{0.50cm}
\centerline{\includegraphics[width=8.50cm]{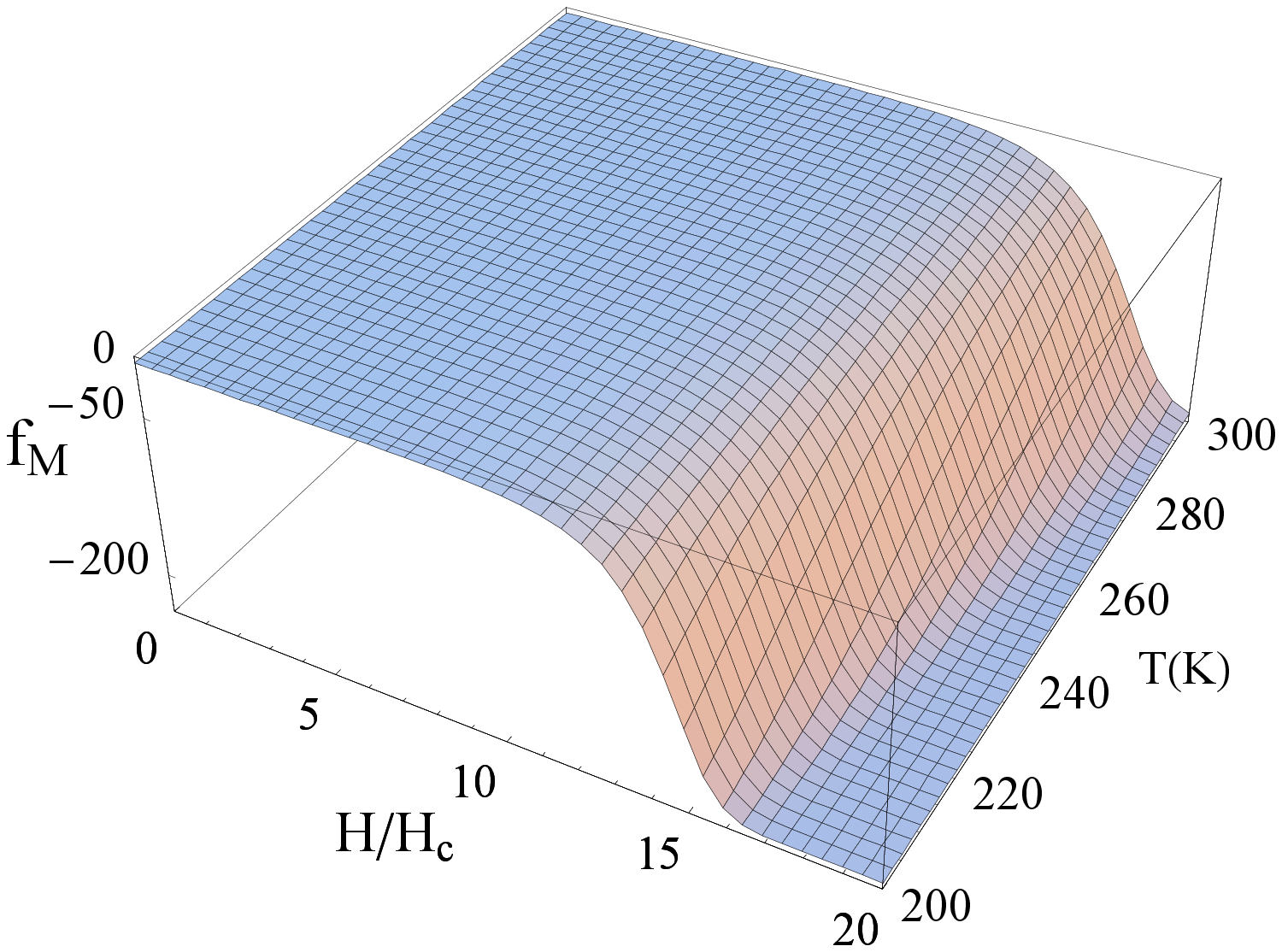}}\vspace{0.1cm}
\caption{$3D$ plots of theoretically predicted hopping $f_h=\Delta R_h(T,H)/R_h(T,0)$ (top) and magnetization governed $f_M=\Delta R_M(T,H)/R_M(T,0)$ (bottom) contributions to magnetoresistance.}
\label{fig:fig5}
\end{figure}
Dependence of the normalized zero-field Curie temperature $t_C(0,x)=T_C(0,x)/T_C(0,x_m)$ and the coercive field $h_c(x)=H_c(x)/H_c(x_m)$ on $x$ is depicted in Fig.4. Notice that both follow percolation laws, namely $T_C(x)\simeq T_C(x_m)(x/x_m)^{\beta}$ and $H_c(x)\simeq H_c(x_m)(x/x_m)^{\gamma}$ with $T_C(x_m)=295K$, $H_c(x_m)=0.33kOe$,  $\beta =0.15$, $\gamma =0.6$, and $x_m=0.2$. In turn, the saturation magnetization is given by $M_s(x_m)\simeq 0.5 \mu_B$ per $Co$ atom ($\mu_B$ is the Bohr magneton). 
According to Fig.3, at high temperatures, the MR $\Delta R_M(T,H)=R_M(T,H)-R_M(T,0)$ is negative because $R_M(T,H)$ \textit{linearly} depends on magnetization $M(T,H)$ which reminds the manifestation of an anomalous Hall effect. \cite{7}  And finally, to extend the validity of our theoretical predictions to relevant temperature intervals, in Fig.5 we present the 3D plots showing the behavior of the MR according to two different magneto-transport scenarios. As we can see, the hopping $\Delta R_h(T,H)$ and magnetization governed $\Delta R_M(T,H)$ contributions could reach as much as $f_h=\Delta R_h(T,H)/R_h(T,0) \simeq +30\%$ and $f_M=\Delta R_M(T,H)/R_M(T,0) \simeq -200\%$, respectively.  

\section{Conclusions}

In conclusion, some very unusual magneto-transport properties of $C_{1-x}Co_x$ thin films (with $x=0.1$, $0.15$ and $0.2$) grown on $Si$ substrate by pulsed laser deposition technique have been reported and attributed to manifestation of magnetic field modulated weak localization scenario (leading to positive magnetoresistance at low temperatures) and a robust electron scattering on ferromagnetic cobalt atoms (leading to negative magnetoresistance at high temperatures).\\

{\bf Acknowledgements}\\

We would like to thank LMA-IQ for allowing us to use FEG-SEM facilities. This work was financially supported by Brazilian agencies CNPq, CAPES, FAPESQ (DCR-PB 302286/2016-0) and FAPESP (CEPID CDMF 2013/07296-2 and 2014/01371-5).

\end{document}